\documentclass[12pt, a4paper]{article}
\usepackage{epsf}
\usepackage{amsmath,amssymb}
\input{colordvi.tex}
\usepackage[usenames,dvipsnames]{color}
\usepackage{graphicx}
\usepackage{cite}
\usepackage[sort&compress, numbers, merge]{natbib}

\usepackage[colorlinks=true, linkcolor=red, citecolor=red,
urlcolor=red]{hyperref}

\begin{document}

\newcommand{\lsim}{\stackrel{<}{_\sim}}
\newcommand{\gsim}{\stackrel{>}{_\sim}}
\renewcommand{\theequation}{\thesection.\arabic{equation}}

\renewcommand{\thefootnote}{\fnsymbol{footnote}}
\setcounter{footnote}{0}

%%%%%%%%%%%%%%%%%%%%%%%%%%%%%%%%%%%%%%%%%%%%%%%%%%

%%%%%%%%%%%%%%%%%%%%%%%%%%%%%%%%%%%%%%%%%%%%%%%%%%
\def\thefootnote{\fnsymbol{footnote}}
\def\a{\alpha}
\def\b{\beta}
\def\c{\varepsilon}
\def\d{\delta}
\def\e{\epsilon}
\def\f{\phi}
\def\g{\gamma}
\def\h{\theta}
\def\k{\kappa}
\def\l{\lambda}
\def\m{\mu}
\def\n{\nu}
\def\p{\psi}
\def\q{\partial}
\def\r{\rho}
\def\s{\sigma}
\def\t{\tau}
\def\u{\upsilon}
\def\v{\varphi}
\def\w{\omega}
\def\x{\xi}
\def\y{\eta}
\def\z{\zeta}
\def\D{\Delta}
\def\G{\Gamma}
\def\H{\Theta}
\def\L{\Lambda}
\def\F{\Phi}
\def\P{\Psi}
\def\S{\Sigma}
\def\me{\mathrm e}

\def\o{\over}
\def\beq{\begin{eqnarray}}
\def\eeq{\end{eqnarray}}
\newcommand{\vev}[1]{ \left\langle {#1} \right\rangle }
\newcommand{\bra}[1]{ \langle {#1} | }
\newcommand{\ket}[1]{ | {#1} \rangle }
\newcommand{\bs}[1]{ {\boldsymbol {#1}} }
\newcommand{\mc}[1]{ {\mathcal {#1}} }
\newcommand{\mb}[1]{ {\mathbb {#1}} }
\newcommand{\EV}{ {\rm eV} }
\newcommand{\KEV}{ {\rm keV} }
\newcommand{\MEV}{ {\rm MeV} }
\newcommand{\GEV}{ {\rm GeV} }
\newcommand{\TEV}{ {\rm TeV} }
\def\diag{\mathop{\rm diag}\nolimits}
\def\Spin{\mathop{\rm Spin}}
\def\SO{\mathop{\rm SO}}
\def\O{\mathop{\rm O}}
\def\SU{\mathop{\rm SU}}
\def\U{\mathop{\rm U}}
\def\Sp{\mathop{\rm Sp}}
\def\SL{\mathop{\rm SL}}
\def\tr{\mathop{\rm tr}}
\def\sp{\;\;}

\def\IJMP{Int.~J.~Mod.~Phys. }
\def\MPL{Mod.~Phys.~Lett. }
\def\NP{Nucl.~Phys. }
\def\PL{Phys.~Lett. }
\def\PR{Phys.~Rev. }
\def\PRL{Phys.~Rev.~Lett. }
\def\PTP{Prog.~Theor.~Phys. }
\def\ZP{Z.~Phys. }
%%%%%%%%%%%%%%%%%%%%%%%%%%%%%%%%%%%%%%%%%%%%%%%%%%

%%%%%%%%%%%%%%%%%%%%%%%%%%%%%%%%%%%%%%%%%%%%%%%%%%
\begin{titlepage}

\begin{center}

\hfill APCTP Pre2016-007\\
\hfill DESY 16-043\\
\hfill UT-16-12\\

\vskip .75in

{\Large \bf 
Structure of K\"{a}hler potential \\ for 
$D$-term inflationary attractor models
}

\vskip .75in

{\large Kazunori Nakayama$^{(a,b)}$, Ken'ichi Saikawa$^{(c,d)}$,\\ Takahiro Terada$^{(a,e)}$,
and Masahide Yamaguchi$^{(d)}$} 

\vskip 0.25in

\begin{tabular}{ll}
$^{a}$&\!\! {\em Department of Physics, Faculty of Science, }\\
& {\em The University of Tokyo,  Bunkyo-ku, Tokyo 133-0033, Japan}\\[.3em]
$^{b}$ &\!\! {\em Kavli IPMU (WPI), UTIAS,}\\
&{\em The University of Tokyo,  Kashiwa, Chiba 277-8583, Japan}\\[.3em]
$^{c}$ &\!\! {\em Deutsches Elektronen-Synchrotron DESY,}\\
&{\em Notkestrasse 85, 22607 Hamburg, Germany}\\[.3em]
$^{d}$ &\!\! {\em Department of Physics, Tokyo Institute of Technology,}\\
&{\em Ookayama, Meguro-ku, Tokyo 152-8551, Japan}\\[.3em]
$^{e}$ &\!\! {\em Asia Pacific Center for Theoretical Physics (APCTP),}\\
&{\em 67 Cheongam-ro, Nam-gu, Pohang 37673, South Korea}\\[.3em]

\end{tabular}

\end{center}
\vskip .5in

\begin{abstract}
Minimal chaotic models of $D$-term inflation predicts too large
primordial tensor perturbations.  Although it can be made
consistent with observations utilizing higher order terms in the
K\"{a}hler potential, expansion is not controlled in the absence of
symmetries.  We comprehensively study the conditions of K\"{a}hler
potential for $D$-term plateau-type potentials and discuss its symmetry.
They include the $\alpha$-attractor model with a massive vector
supermultiplet and its generalization leading to pole inflation of
arbitrary order.  We extend the models so that it can describe Coulomb
phase, gauge anomaly is cancelled, and fields other than inflaton are
stabilized during inflation.  We also point out a generic issue for
large-field $D$-term inflation that the masses of the non-inflaton fields
tend to exceed the Planck scale.
\end{abstract}

\end{titlepage}

\renewcommand{\thepage}{\arabic{page}}
\setcounter{page}{1}
\renewcommand{\thefootnote}{\#\arabic{footnote}}
\setcounter{footnote}{0}
\baselineskip 0.58cm
%%%%%%%%%%%%%%%%%%%%%%%%%%%%%%%%%%%%%%%%%%%%%%%%%%

%%%%%%%%%%%%%%%%%%%%%%%%%%%%%%%%%%%%%%%%%%%%%%%%%
\section{Introduction}
%%%%%%%%%%%%%%%%%%%%%%%%%%%%%%%%%%%%%%%%%%%%%%%%%

Inflation in the early universe is not only indispensable to explain
horizon and longevity problems (see recent review,
e.g. \cite{Sato:2015dga}) but also is strongly supported by the
observations of cosmic microwave background anisotropies
\cite{Bennett:2012zja,Hinshaw:2012aka,Ade:2015xua,Ade:2015lrj}. The
position of the first peak of the temperature-temperature correlations
suggests spatially flat Universe as predicted by inflation, and the
large scale anti-correlations in the temperature-E mode correlations
implies the superhorizon epoch of primordial curvature perturbations,
which can be naturally realized by inflation. Unfortunately, primordial
tensor perturbations have not yet been found, which strongly constrains
inflation model building. Actually, the recent results of the PLANCK
collaboration place constraint on the tensor-to-scalar ratio as $r <
0.10$ (95\% CL, Planck TT+lowP) at $k = 0.002 {\rm Mpc}^{-1}$, and when
combined with BICEP2/Keck-Array, $r < 0.08$ (95\% CL, Planck
TT+lowP+BKP) at the same pivot scale~\cite{Ade:2015xua,
Ade:2015lrj}. This almost ruled out chaotic inflation models with
quadratic and higher power-law potential. Instead, cosmological
attractor models
\cite{Kallosh:2013tua,Pallis:2013yda,Giudice:2014toa,Pallis:2014dma,Kallosh:2014rga,Kallosh:2014laa,Galante:2014ifa,Carrasco:2015uma}
including conformal attractor model \cite{Kallosh:2013hoa} and its
extension to $\alpha$-attractor models \cite{Ellis:2013nxa,
Ferrara:2013rsa,Kallosh:2013yoa,Carrasco:2015pla} have recently been
getting more attention because they can easily fit the observed spectral
index of the primordial curvature perturbations and accommodate
observable tensor-to-scalar ratio rather arbitrarily. In
Ref. \cite{Galante:2014ifa}, it became manifest that pole structure of
order 2 in a kinetic term is important to realize this kind of
$\alpha$-attractor models. This idea is further extended to pole
inflation\footnote{This terminology should not be confused with the
pole-law inflation \cite{Pollock:1989vn} proposed decades ago.}
\cite{Galante:2014ifa, Broy:2015qna, Terada:2016nqg}, in which, around a pole of arbitrary order in the kinetic term, the
potential becomes effectively flat after canonical normalization of the
kinetic term.

On the other hand, in order to realize inflation including the above
attractor models, supersymmetry (SUSY) is desired, which can control
radiative corrections to the inflaton field to preserve the flatness of
its potential. However, if one introduces local SUSY or supergravity,\footnote{
For recent reviews of inflation in supergravity, see Refs.~\cite{Mazumdar:2010sa, Yamaguchi:2011kg, Terada:2015sna}
} we
encounter difficulties again as corrections of order of the Hubble scale
to the inflaton mass in $F$-term inflation models
\cite{Copeland:1994vg}. This is because positive vacuum energy during
inflation breaks SUSY and its effect as a SUSY breaking mass is in general 
transmitted to scalar fields including the inflaton. One way
to circumvent this difficulty is to introduce a symmetry such as a shift
symmetry \cite{Kawasaki:2000yn, *Kawasaki:2000ws} to protect the flatness
of the potential if one does not wish to adopt ad-hoc K\"ahler
potential \cite{Murayama:1993xu}. Another interesting possibility is to
consider $D$-term inflation \cite{Binetruy:1996xj,Halyo:1996pp}, which
is free from the above problem thanks to the absence of the Hubble scale
correction. The original $D$-term inflation was proposed as hybrid
inflation and later was extended to chaotic inflation
\cite{Kadota:2007nc, Kawano:2007gg, Kadota:2008pm}. However, as the
observations are getting more and more precise, the predictions of
simple models of $D$-term inflation tend to conflict with the
observational data unfortunately. Of course, as explicitly given in
Sec.~\ref{sec:D}, if we take higher order terms of K\"ahler potential
into account, such $D$-term inflation models can still fit the
observational data. But, such higher order terms are uncontrollable
without symmetry.  Then, instead, we revisit the attractor models in the
context of $D$-term inflation, which were discussed in the context of a
massive vector supermultiplet~\cite{Ferrara:2013rsa,Ferrara:2013kca} and
a Dirac-Born-Infeld action~\cite{Abe:2015fha}. In this paper, we derive
general conditions of K\"{a}hler potential necessary to realize $D$-term
inflationary attractor models in the context of pole inflation.  Along
the way, we point out a generic issue in large-field $D$-term inflation
to stabilize additional matter fields without knowledge of quantum
gravity.

The paper is organized as follows. In the next section, we briefly
review $D$-term inflation models in supergravity and discuss the
conflict of their predictions of simple models with the current data. As
a simple extension, we take into account higher order corrections in
K\"ahler potential and show how well such corrections improve the fit to
the data. In Sec.~\ref{sec:alpha}, we revisit the
$\alpha$-attractor models for $D$-term inflation and extend the theory
to the symmetric phase of gauge symmetry as well. We give a
concrete workable example, in which the theory is well-defined almost
everywhere on the field space and fields required for anomaly
cancellation are stabilized during inflation. We also elucidate
symmetries guaranteeing the (approximate) flatness of the $D$-term
potential. 
In Sec.~\ref{sec:pole}, we generalize these results and derive a generic
condition for K\"ahler potential to realize $D$-term inflationary
attractor models in the context of pole inflation.  Final section is
devoted to conclusion and discussions.

%%%%%%%%%%%%%%%%%%%%%%%%%%%%%%%%%%%%%%%%%%%%%%%%%%
\section{Large field $D$-term inflation}
\label{sec:D}
\setcounter{equation}{0}
%%%%%%%%%%%%%%%%%%%%%%%%%%%%%%%%%%%%%%%%%%%%%%%%%%

In this section we briefly review large field $D$-term inflation models
and their problems.  Hereafter, we take the reduced Planck unit $M_P=1$
unless otherwise stated.  For the K\"ahler potential
$K(\Phi_i,\Phi^*_{\bar j})$, the superpotential $W(\Phi_i)$ and the
gauge kinetic function $f(\Phi_i)$, the supergravity Lagrangian for the
scalar field $\{ \Phi_i \}$ in the Einstein frame is
\begin{align}
	\mathcal L = K_{i\bar j}(\partial^{\mu} \Phi_i)(\partial_{\mu} \Phi_{\bar j}^*) - V.  \label{Lag}
\end{align}
The scalar potential consists of $F$-term and $D$-term potential,
\begin{align}
	&V=V_F + V_D,\\
	&V_F = e^K \left[ K^{i\bar j}(D_iW)(D_{\bar j} \bar W) - 3|W|^2 \right], \label{VF}\\
	&V_D = \frac{1}{2\,{\rm Re} f}\left( i K_i X_i \right)^2,  \label{VD}
\end{align}
where $K^{i\bar j} = K_{i\bar j}^{-1}$ is the inverse matrix of the K\"{a}hler metric $K_{i\bar{j}}$, $D_i W = W_i + K_i W$ is the K\"{a}hler covariant derivative, and $X_i$ is the Killing vector of the K\"{a}hler manifold, which is $X_i=i q_i \Phi_i$ in the case of linearly transforming fields under the U(1) symmetry with $q_i$ being the U(1) charge of $\Phi_i$. 
In this case, the $D$-term potential reduces to the usual form,
\begin{align}
V_D = \frac{1}{2\,{\rm Re} f}\left(\sum_i q_i K_i \Phi_i \right)^2. \label{VD_2}
\end{align}
Here, we do not consider the possibility of field-independent
Fayet-Iliopoulos (FI) term or gauged U(1) $R$-symmetry for simplicity.

%%%%%%%%%%%%%%%%%%%%%%%%%%%%%%%%%%%%%%%%%%%%%%%%%%
\subsection{$D$-term inflation with monomial/polynomial potential}
%%%%%%%%%%%%%%%%%%%%%%%%%%%%%%%%%%%%%%%%%%%%%%%%%%

For simplicity, we consider a setup with two superfields $\Phi_+$ and
$\Phi_-$ charged under U(1) gauge symmetry with charge $+1$ and $-1$,
respectively.  Note that this is a minimal setup to avoid the gauge
anomaly.  Later we regard $\varphi \equiv \sqrt{2}|\Phi_+|$ as an
inflaton.\footnote{All the following discussions do not change if we
regard $\Phi_-$, instead of $\Phi_+$, as an inflaton.} 
We also introduce a gauge-singlet superfield $S$ to stabilize fields other than
the inflaton.\footnote{Note that it may be possible that there is a non-trivial inflationary trajectory in a field space spanned by $\Phi_+$ and $\Phi_-$
	without introducing any other stabilization term. 
	For example, inflation can happen while $\Phi_+$ and $\Phi_-$ are rolling down toward the $D$-flat direction.
	In this paper, we want to restrict ourselves to the case of exact single-field inflation as a first step to the model building,
	in which either $|\Phi_+|$ or $|\Phi_-|$ is responsible for inflation and other fields are heavy enough to be integrated out.}
In the case of $F$-term models with stabilizer superfield
$S$ whose $F$-term component breaks SUSY during inflation, we can
introduce a higher dimensional term like $K \sim - |\Phi|^2 |S|^2$ where
$\Phi$ is a field we want to stabilize during inflation.  This generates
a SUSY breaking mass term for $\Phi$.  In our case of $D$-term SUSY
breaking during inflation, it is difficult to emulate this mechanism
because the superfield acquiring $D$-term is the gauge (vector)
superfield, whose interaction is completely determined by the gauge
symmetry.  After all, this kind of coupling is just a part of $D$-term
potential.  Therefore, we introduce a superpotential to stabilize
additional fields.  We assume the superpotential of the form
\begin{align}
	W= \lambda S \Phi_+ \Phi_-.  \label{W}
\end{align}
This can be viewed as a $\Phi_+$-dependent mass term for $S$ and $\Phi_-$.
Then, for $S=0$ and $\Phi_-=0$, the $F$-term potential is exactly flat for any $|\Phi_+| > 0$.
Hereafter we consider this configuration.
Stability against $S$ and $\Phi_-$ directions will be discussed in Sec.~\ref{sec:stab}.
We also take $f=1/g^2$, with $g$ being a gauge coupling constant, in the following.

First of all, let us consider the minimal K\"ahler potential:
\begin{align}
	K = |\Phi_+|^2 + |\Phi_-|^2 + |S|^2.
\end{align}
The kinetic term is canonical and the $D$-term potential for $\Phi_-=0$ is given by \cite{Kadota:2008pm}
\begin{align}
	V = \frac{g^2 \varphi^4}{8}.
\end{align}
Therefore, we obtain a simple quartic inflaton potential and inflation occurs for $\varphi \gg 1$.\footnote{The angular component of $\Phi_+$ is a Goldstone mode and does not affect the inflaton dynamics.}
This is a simple realization of $\varphi^4$ model for chaotic inflation, but it contradicts with observations~\cite{Ade:2015lrj}
since it predicts too large tensor-to-scalar ratio $r = 16/N_e$ with $N_e$ being the $e$-folding number 
after the observable scale exits the horizon. Typically we have $N_e \simeq 50- 60$ depending on the thermal history after inflation.

One may extend the minimal model to include higher order terms in the K\"ahler potential:
\begin{align}
	K = \sum_n \frac{k_n}{n} \left( |\Phi_+|^{2n} + |\Phi_-|^{2n}+|S|^{2n} \right).   \label{K_model1}
\end{align}
We take $k_1=1$ to make $\Phi_i$ ($i=+, -,S$) canonically normalized in the small field limit $\Phi_i \to 0$.
The $D$-term potential for $\Phi_+$ with $\Phi_-=0$ is given by
\begin{align}
	V = \frac{g^2}{2}\left( \sum_n k_n |\Phi_+|^{2n} \right)^2.
\end{align}
To be concrete, we assume the hierarchy $1 \gg |k_2| \gg |k_3| \gg \dots$ and neglect terms with $k_n$ $(n\geq 4)$.
Then the kinetic term is given by
\begin{align}
	\mathcal L_{\text{K}} = \frac{1}{2}\left(1+ k_2 \varphi^2 + \frac{3k_3}{4} \varphi^4\right)(\partial_{\mu} \varphi)^2, \label{LK_model1}
\end{align}
and the $D$-term potential is given by
\begin{align}
	V = \frac{g^2\varphi^4}{8}\left(1+\frac{k_2}{2}\varphi^2 + \frac{k_3}{4}\varphi^4 \right)^2. \label{V_model1}
\end{align}
As shown in Refs.~\cite{Destri:2007pv,Nakayama:2013jka, *Nakayama:2013txa}, this kind of modification of the inflaton potential
from the monomial to polynomial significantly changes the prediction of $(n_{\text{s}},r)$ and we can make the fit to the observation better. 
In contrast to Refs.~\cite{Destri:2007pv,Nakayama:2013jka, *Nakayama:2013txa}, 
 both the kinetic term~\eqref{LK_model1} and potential~\eqref{V_model1} are determined solely by the K\"ahler potential
and hence the structure of the potential in the canonical basis is more complicated.
We assume $k_2 < 0$ in the following to improve the fit to observation.
Note that we need the following condition if we require that $K_{\Phi_+ \bar\Phi_+} > 0$ for all $\varphi>0$,
\begin{align}
	k_3 >\frac{1}{3}k_2^2.
\end{align}
We can calculate the spectral index and tensor-to-scalar ratio by using the method given in Appendix~\ref{sec:app}.
The result is shown in Fig.~\ref{fig:model1}.
By choosing the parameter, we can make the prediction consistent with the Planck observation~\cite{Ade:2015lrj}.
The overall magnitude of the density perturbation is reproduced for $g\sim 10^{-6}$, 
hence this U(1) group cannot be identified with the subgroup of the standard model gauge group.

\begin{figure}[t]
\begin{center}
\includegraphics[width=0.618034 \columnwidth]{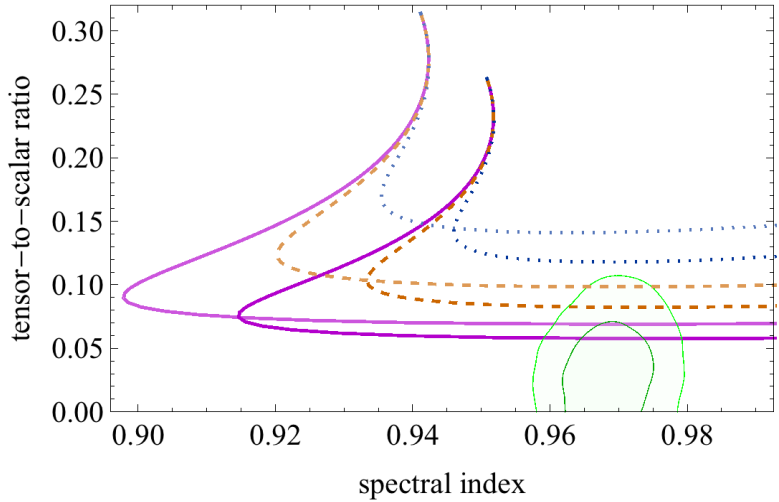} 
\caption{
	$(n_s,r)$ for $d=0.35$ (purple solid), 0.36 (orange dashed), and 0.38 (blue dotted), where we have assumed $d=k_3/k_2^2$. 
	Lighter and darker lines correspond to the cases with $N_e = 50$ and $60$, respectively.	
	For each line, we varied $-k_2=10^{-5}\textendash 3\times 10^{-3}$.
	Green curves represent 68\% and 95\% confidence regions of the Planck TT+lowP+BKP+lensing+BAO+JLA+$H_0$ constraint (adopted from Fig.~21 of Ref.~\cite{Ade:2015xua}).
}
\label{fig:model1}
\end{center}
\end{figure}

Another model is given in Appendix~\ref{app:generic}, in which K\"{a}hler potential is expanded in terms of logarithm of inflaton field rather than the inflaton field itself.
Due to this change, expansion of the canonical inflaton potential starts with lower powers than the quartic.
Similarly to the above example, it is possible for its prediction to lie within the Planck contour by utilizing higher order terms in the K\"{a}hler potential. 

Although it is possible to construct a model consistent with observations as above, it has several drawbacks.
First, the K\"ahler potential is not controlled by any symmetry,
hence there is no reason to expect that higher-order terms do not make comparable contributions to the results,
which poses some doubts about the predictability of the model.
Second, our calculation so far is based on an assumption that $S$ and $\Phi_-$ are stabilized at $S=\Phi_-=0$.
However, this assumption is not always justified, as we will see below.

%%%%%%%%%%%%%%%%%%%%%%%%%%%%%%%%%%%%%%%%%%%%%%%%%%
\subsection{Stabilization of other fields}  \label{sec:stab}
%%%%%%%%%%%%%%%%%%%%%%%%%%%%%%%%%%%%%%%%%%%%%%%%%%

Because of the structure of the $D$-term potential~\eqref{VD_2}, the
oppositely charged field $\Phi_-$ has a tachyonic mass during inflation,
 \begin{align}
 m^2_{\Phi_{-}, D} = - \sqrt{6} g H,
 \end{align}
where the subscript $D$ ($F$) represents the fact that these masses are
obtained from $D$-term ($F$-term) potential, and $H$ denotes the Hubble
parameter.  In order to avoid such a tachyonic instability, we need to
introduce the $F$-term potential, which gives positive masses squared
for $\Phi_-$ and $S$.  With the superpotential \eqref{W}, kinetic terms
and $F$-term potential for $S$ and $\Phi_-$ during inflation are given
by
\begin{align}
	&\mathcal L_{\text{K}} = K_{S\bar S}\left|\partial_{\mu} S \right|^2 + K_{\Phi_-\bar \Phi_-}\left|\partial_{\mu} \Phi_-\right|^2,\\
	&V_F = e^K \frac{\lambda \varphi^2}{2}\left[ K_{S\bar S}^{-1} |\Phi_-|^2 + K_{\Phi_-\bar \Phi_-}^{-1} |S|^2  \right],
\end{align}
where we have assumed that K\"ahler metric is diagonal around $S=\Phi_-=0$.
This scalar potential to stabilize $S$ and $\Phi_-$ is actually so steep
that their masses easily exceed the Planck scale due to the large
exponential factor $e^{K} \simeq e^{{\cal O}(N_e)}$.
If we demand that their masses are larger than the Hubble scale at the end of inflation, their masses at the $e$-folding $60$ is
exponentially larger than the Hubble scale during inflation, which is
typically much larger than the Planck scale. It means that quantum
gravity effects may become important for inflationary region and the
validity of the calculation is lost.  This may be a general feature of
large field $D$-term inflation models in which fields other than the
inflaton is strongly stabilized during inflation.\footnote{Note that all
the mass scales in the $F$-term potential, including minimal SUSY
standard model sector and SUSY breaking sector, are exponentially
enhanced during inflation. While it is possible that there are no mass
scales during inflation except for the $D$-term potential in the
inflaton sector, we stress that it might be non-trivial to have
successful reheating and realistic SUSY model in such a setup.}

This problem may be avoided by extending the structure of the K\"ahler potential for $S$ and $\Phi_-$.
For example, let us take the following K\"ahler potential:
\begin{align}
	K = \sum_n \frac{k_n}{n} \left( |\Phi_+|^{2n} + |\Phi_-|^{2n} +|S|^{2n} \right) +h |S|^2 \left(|\Phi_+|^2 e^{\sum_n \frac{k_n}{n} |\Phi_+|^{2n} } + |\Phi_-|^2 e^{\sum_n \frac{k_n}{n} |\Phi_-|^{2n} }\right),
\end{align}
where $h$ is a constant.  The addition of the last two terms does
not affect the inflaton potential at $\Phi_- = S=0$.  Then the physical
$F$-term masses of $S$ and $\Phi_-$ in the canonical basis are
\begin{align}
	m_{\Phi_-,F}^2 = m_{S,F}^2 = e^K K_{S\bar S}^{-1} K_{\Phi_-\bar \Phi_-}^{-1}\frac{\lambda^2 \varphi^2}{2} \simeq \frac{\lambda^2}{h}.
	\label{mass_F_model1}
\end{align}
Note that $m_{\Phi_-,F}^2$ and $m_{S,F}^2$ are almost identical during inflation since we can replace the factor $K^{i\bar{i}}$ in Eq.~\eqref{VF}
with $K_{i\bar{i}}^{-1}$ for $\Phi_-=S=0$.
Therefore large exponential dependence is cancelled out with this kind of ad-hoc assumption on the K\"ahler potential
and hence we can avoid the problem of too large masses for these fields.
On the other hand, $D$-term masses for $S$ and $\Phi_-$ are given by
\begin{align}
m_{\Phi_-,D}^2 = -\sqrt{6}g H \quad\text{and}\quad & m_{S,D}^2 = 6H^2.
\end{align}
Therefore, we need $m_{\Phi_-,F}^2 + m_{\Phi_-,D}^2 \gtrsim H^2$, which implies
$\lambda^2 \gtrsim h g H$, in order to stabilize $\Phi_-$ at the origin during inflation.

%%%%%%%%%%%%%%%%%%%%%%%%%%%%%%%%%%%%%%%%%%%%%%%%%%
\section{Revisiting $D$-term $\alpha$-attractor model}
\label{sec:alpha}
\setcounter{equation}{0}
%%%%%%%%%%%%%%%%%%%%%%%%%%%%%%%%%%%%%%%%%%%%%%%%%%

\subsection{$D$-term $\alpha$-attractor demystified and its symmetry}

The $D$-term $\alpha$-attractor model~\cite{Ferrara:2013rsa} had been
constructed before the ($F$-term) superconformal $\alpha$-attractor was
discovered~\cite{Kallosh:2013yoa}, but subsequent discussions of
inflationary attractor models were mainly on $F$-term models.  Later,
attractor models were understood as theories with second order pole in
the kinetic term~\cite{Galante:2014ifa}, which was further extended to
arbitrary order poles~\cite{Galante:2014ifa, Broy:2015qna,
Terada:2016nqg}.  So, let us tentatively pretend as if we do not know
about the original $D$-term attractor model, and consider its
possibility in the perspective of general attractor or pole inflation
context.

As repeatedly mentioned, the essence of the $\alpha$-attractor or more generally pole inflation is the presence of a pole in the kinetic term of the inflaton field before canonical normalization.
With this in mind, a naive gauge-invariant choice for a possible $D$-term attractor-type model would be
\begin{align}
K = -3 \alpha \log \left( 1 - |\Phi|^2 \right),
\end{align}
where $\Phi$ is the inflaton field charged under U(1) symmetry, for simplicity, and other possible charged fields are neglected at this stage.
This K\"{a}hler manifold has a curvature 
\begin{align}
R\equiv & \left(K^{\bar{\Phi}\Phi}\right)^2 
\left( K^{\bar{\Phi}\Phi}K_{\Phi\bar{\Phi}\bar{\Phi}}K_{\Phi\bar{\Phi}\Phi} - K_{\Phi\bar{\Phi}\Phi\bar{\Phi}}\right)  \label{curvature}\\
=& -\frac{2}{3\alpha},
\end{align}
 and leads to a second order pole in the kinetic term at $|\Phi|=1$,
\begin{align}
K_{\Phi\bar{\Phi}}=\frac{3\alpha}{(1-|\Phi|^2)^2}.
\end{align}
This apparently perfectly matches the requirement of pole inflation, and
indeed this K\"{a}hler potential is used for $F$-term $\alpha$-attractor
models~\cite{Kallosh:2013yoa}, though the appearance of the pole in the
factor $e^K$ of scalar potential is cancelled out by the suitable choice
of superpotential. There is another kind of difficulty in $D$-term cases
because not only the kinetic term but also the $D$-term potential is
determined by K\"{a}hler potential.  When the K\"{a}hler metric $K_{\Phi
\bar{\Phi}}$ has a second order pole, the $D$-term potential, which
contains $|K_{\Phi}|^2$, also has a second order pole generically.  For
the above K\"{a}hler potential, the scalar potential is
\begin{align}
V&= \frac{g^2}{2} \left( \frac{ 3 \alpha |\Phi|^2}{1- |\Phi|^2} \right)^2 \nonumber \\
&= \frac{g^2}{2} (3\alpha )^2 \sinh^{4} \left( \frac{\tilde{\varphi}}{\sqrt{6\alpha}}\right),
\end{align}
where $\tilde{\varphi}=\sqrt{6\alpha} \tanh^{-1}(|\Phi|)$ is the
canonical inflaton field.  Thus, after canonical normalization, we
obtain an exponentially steep function.  Although a suitable choice of
gauge kinetic function might in general rescue this situation like
superpotential in $F$-term cases, it is unlikely in our minimal setup
because SUSY and gauge invariance require $f(\Phi_{+},\,
\Phi_{-})=f(\Phi_{+}\Phi_{-})$, and this reduces to a constant in our
configuration $\Phi_{-}=0$.  Exponential stretching due to canonical
normalization is still valid, but the value of the potential grows
rapidly near the pole, so the exponential flattening is not realized.
This is because of the presence of the second order pole in the original
scalar potential~\cite{Terada:2016nqg}.

The $\alpha$-attractor model in the $D$-term case was first discussed as
an example of massive vector (or tensor) supermultiplet
models~\cite{Ferrara:2013rsa}.  In this approach, a U(1) vector
supermultiplet and a chiral supermultiplet corresponding to a
Higgs/Nambu-Goldstone field are introduced.  Instead of taking
Wess-Zumino gauge, gauge is chosen so that the chiral multiplet becomes
trivial leading to the theory of a massive vector multiplet.  This is
just a SUSY description of the Higgs mechanism.

Here, we take Wess-Zumino gauge as usual, and retain a chiral superfield.
Then, the $D$-term model corresponding to the $\alpha$-attractor potential is given by~\cite{Ferrara:2013rsa}\footnote{
Equation~(4.31) of Ref.~\cite{Ferrara:2013rsa} contains a typo: either the term inside the logarithm or the second term should be multiplied by minus one.
},
\begin{align}
K = -3 \alpha \log \left( \Lambda + \bar{\Lambda} \right) + 3\beta \left( \Lambda + \bar{\Lambda} \right) ,\label{K_shift}
\end{align}
where $\alpha$, $\beta>0$.
$\Lambda$ transforms nonlinearly like axion under the U(1) transformation, \textit{i.e.} $\Lambda \to \Lambda'=\Lambda - c X_{\Lambda}$ where $c$ is the transformation parameter, and the Killing vector is now given by $X_{\Lambda}= i q$ with some constant $q$. 
The $\Lambda$ field may be related to the usual U(1) charged field $\Phi$ as $\Phi= e^{\Lambda}$, and $q$ is interpreted as the U(1) charge of $\Phi$.
Inflation occurs in the large-field region of $\Lambda+\bar{\Lambda}$, and the imaginary part is absorbed by the gauge field due to the Higgs mechanism.
It should be noted that the inflaton is not the shift symmetric direction in this case in contrast to the standard lore~\cite{Kawasaki:2000yn, *Kawasaki:2000ws, Kallosh:2010ug, Kallosh:2010xz}\footnote{
It is possible to rewrite Eq.~\eqref{K_shift} in such a way that shift symmetry of the superfield $\phi=\log(\Lambda )$ corresponds to the shift symmetry of the canonical inflaton up to K\"{a}hler transformation,
\begin{align}
K = - \frac{3}{2} \alpha \log \left(2 +  e^{\phi -\bar{\phi}}+ e^{-(\phi - \bar{\phi})} \right) -\frac{3}{2}\alpha (\phi + \bar{\phi}) + 3\beta (e^{\phi}+e^{\bar{\phi}}).
\end{align}
In this expression, each of the first two terms is not gauge invariant, but it is when added up.
}.
Note that the shift symmetry of $\Lambda$ in Eq.~\eqref{K_shift} becomes U(1) symmetry of $\Phi$ in the K\"{a}hler potential
\begin{align}
K= -3 \alpha \log \left( +\log \left( |\Phi|^2 \right) \right) + 3 \beta \log \left( |\Phi|^2 \right). \label{K_phi_1-inf}
\end{align}
This kind of change of variables was discussed in the context of $F$-term inflation~\cite{Li:2014unh}.
In this parametrization of the field space, $\Phi$ is defined only in $|\Phi|>1$.
Inflation happens at large $|\Phi|$, and the phase component is absorbed by the Higgs mechanism.
According to the general discussion of pole inflation~\cite{Galante:2014ifa, Broy:2015qna}, which includes $\alpha$-attractor, potential can be flattened by a pole of the coefficient of kinetic term.
To compare it with this notion, it may be useful to redefine the field $\Phi \to \tilde{\Phi}= 1/\Phi$,
\begin{align}
K= -3 \alpha \log \left( -\log \left( | \tilde{\Phi} |^2 \right) \right)  - 3 \beta \log \left( | \tilde{\Phi}  |^2 \right). \label{K_phi_0-1}
\end{align}
In this parametrization, $\tilde{\Phi}$ is defined only in $0<|\tilde{\Phi}|<1$, and inflation occurs near $\tilde{\Phi}=0$.

Let us see how this model circumvents the issue discussed above.
The K\"{a}hler metric following from Eq.~\eqref{K_phi_0-1} has a second order pole multiplied by logarithmic singularity, so the kinetic term is
\begin{align}
\mathcal{L}_{\text{K}}=\frac{3\alpha}{|\tilde{\Phi}|^2 \left( \log |\tilde{\Phi}|^2 \right)^2} \left| \partial_{\mu}\tilde{\Phi} \right|^2 \left(=\frac{3\alpha}{|\Phi|^2 \left( \log |\Phi|^2 \right)^2} \left| \partial_{\mu}\Phi  \right|^2 \right).  \label{kin_pole}
\end{align}
Actually, the first derivative of the K\"{a}hler potential has a pole of first order, but it is cancelled by the Killing vector $X_i$ [\textit{cf.}~Eq.~\eqref{VD}] because the pole is at the origin,
\begin{align}
V= \frac{g^2}{2} \left( 3\beta + \frac{3\alpha}{\log|\tilde{\Phi}|^2} \right)^2  \left(= \frac{g^2}{2} \left( 3\beta - \frac{3\alpha}{\log|\Phi|^2} \right)^2 \right) .  \label{pot_pole}
\end{align}
Of course, we cannot simply take $K=-3\alpha \log|\Phi|^2$.  Although the first order pole is cancelled similarly, the K\"{a}hler metric vanishes.

The canonical inflaton is 
\begin{align}
\tilde{\varphi}=\sqrt{\frac{3 \alpha}{2}} \log \left( \frac{\beta}{\alpha} \log |\Phi|^2\right),
\end{align}
and the potential becomes
\begin{align}
	V = \frac{9g^2\beta^2}{2}\left[1-\exp\left(-\sqrt{\frac{2}{3\alpha}}\tilde\varphi \right) \right]^2. \label{Vatt}
\end{align}
As is clear from this potential, for $\alpha \to \infty$ this approaches to the quadratic chaotic inflation model.
For $\alpha \lesssim 1$, on the other hand, this shows the attractor behavior:
\begin{align}
	n_s = 1-\frac{2}{N_e},~~~~r=\frac{12\alpha}{N_e^2}.  \label{att_ns-r}
\end{align}

We clarify an approximate global symmetry hidden in the K\"{a}hler potential~\eqref{K_phi_1-inf} or~\eqref{K_phi_0-1}. 
First, note that, in the plateau-type potentials, which asymptotes to a constant in the large field limit, canonical inflaton $\phi$ has an approximate shift symmetry $\phi \to \phi'= \phi - \tilde{c}$ where $\tilde{c}$ is a transformation parameter.
In the case of $D$-term attractor, 
$\phi=  \sqrt{3\alpha /2}\log [(\Lambda + \bar{\Lambda})/\sqrt{2}]$, 
so the shift symmetry of the canonical inflaton corresponds to scale symmetry of $\Lambda = \log \Phi$, \textit{i.e.} $\Lambda \to \Lambda' = e^{-\sqrt{2/3\alpha}\tilde{c}} \Lambda$.  In this respect of scale symmetry in $\alpha$-attractor, see Ref.~\cite{Carrasco:2015uma}.
This in turn implies that the inflaton part of the theory has a symmetry under the ``power transformation'' 
\begin{align}
\Phi \to \Phi' = \Phi^{1/\hat{c}}, \label{powerT}
\end{align}
where $\hat{c}\equiv e^{\sqrt{2/3\alpha}\tilde{c}}$.
This symmetry is respected by the inflaton kinetic and potential terms, but not necessarily respected by other fields. 
By the power transformation \eqref{powerT}, the form of the K\"{a}hler potential \eqref{K_phi_1-inf} changes as
\begin{align}
K=& -3 \alpha \log \left( +\log \left( |\Phi|^2 \right) \right) + 3 \beta \log \left(|\Phi|^2 \right) \nonumber \\
=& -3 \alpha \log \left( +\log \left( |\Phi'|^2 \right) \right) -3 \alpha \log \hat{c}+ 3 \hat{c} \beta \log \left( |\Phi'|^2 \right) .
\end{align}
That is, this is a K\"{a}hler transformation and rescaling of $\beta$.
The constant K\"{a}hler transformation does not affect the inflaton Lagrangian.
Let us separately consider the inflaton kinetic term and potential.
The kinetic term depends only on the term proportional to $\alpha$ because $\log|\Phi|^2=\log\Phi + \log\Phi^*$ is a sum of (anti)holomorphic terms.
Thus, the kinetic term is invariant.
The dominant contribution to the inflaton potential comes from the term proportional to $\beta$.
The rescaling of $\beta$ is compensated by the rescaling of U(1) charge of the inflaton due to Eq.~\eqref{powerT}.
Thus, the dominant (constant) term in the potential is invariant.
Alternatively, one can define the accompanying transformation of vector superfield $V\to V'=V/\hat{c}$, so that the gauge invariant combination transforms covariantly like $(\Phi^*e^{V}\Phi) = (\Phi'{}^{*}e^{V'}\Phi')^{\hat{c}}$.
In this case, the U(1) charge of the inflaton is not changed, but the normalization of gauge kinetic term changes as $\frac{1}{g^2}\mathcal{W}\mathcal{W} =\frac{ \hat{c}^2}{g^2}\mathcal{W}'\mathcal{W}'$.
Because of this rescaling of the gauge coupling (combined with the rescaling of $\beta$ in the K\"{a}hler potential), the resultant $D$-term is invariant.
The term proportional to $\alpha$ breaks the symmetry softly to make a non-trivial (non-constant) inflationary potential.

As a final remark on the $D$-term $\alpha$-attractor model, we comment on the connection to FI term.
Explicitly writing the vector superfield, we can rewrite the K\"{a}hler potential~\eqref{K_phi_1-inf} as
\begin{align}
K= -3 \alpha \log\left[\log\left(\Phi^* e^{V}\Phi\right)\right] + 3\beta \log |\Phi|^2+ 3 \beta V .
\end{align}
As the last term implies, it is as if we have an FI term proportional to $\beta$.
This explains why we can have a constant in the $D$-term potential.
However, it is accompanied by the $\log |\Phi|^2$ term, and the K\"{a}hler potential itself is gauge invariant.
This combination does not cause the problem of gauge non-invariant supercurrent supermultiplet, which is characteristic of an FI term~\cite{Komargodski:2009pc, *Komargodski:2010rb} (see also Ref.~\cite{Dienes:2009td}).
Another way to see similarity to FI term is to apply K\"{a}hler transformation.
Suppose there is a nonzero superpotential.
By K\"{a}hler transformation, we can transfer the second term in \eqref{K_phi_1-inf} into the superpotential,
\begin{align}
K \to  K'= -3 \alpha \log\left[\log\left(\Phi^* e^{V}\Phi\right)\right] + 3 \beta V \quad \text{and} \quad  W \to  W'= \Phi^{3\beta}W.
\end{align}
This is the same form as a gauged U(1) $R$-symmetric theory with a field-independent FI term proportional to $\beta$.
However, as discussed \textit{e.g.}~in Refs.~\cite{Binetruy:2004hh, Catino:2011mu}, this is an imposter of field-independent FI term due to the K\"{a}hler transformation, because the original K\"{a}hler potential and superpotential are separately gauge invariant and the constant term is originated from the $K_i X_i$ term in the $D$-term potential.
In this sense, this is classified as a ``field-dependent'' FI term although it is actually constant (field-independent).
On the other hand, the exponential term in the canonical inflaton potential is originated from the term dependent on $\alpha$ in the K\"{a}hler potential, which has a very similar form to the case of the standard field-dependent FI term~\cite{Dine:1987xk}.
Usually, the modulus field on which the FI term depends is supposed to be stabilized by some mechanism to make the FI term effectively constant (but it is a nontrivial task, see \textit{e.g.}~Refs.~\cite{Binetruy:2004hh, Wieck:2014xxa}).
In the case of $D$-term attractor model, such stabilization is no longer required because there is a truly constant term separately and the modulus is identified as the inflaton which slowly rolls down the potential.

%%%%%%%%%%%%%%%%%%%%%%%%%%%%%%%%%%%%%%%%%%%%%%%%%%
\subsection{Modification of $D$-term $\alpha$-attractor model}
%%%%%%%%%%%%%%%%%%%%%%%%%%%%%%%%%%%%%%%%%%%%%%%%%%

If we identify $\Phi$ as a usual chiral superfield charged under U(1) symmetry, we want to be able to describe not only the Higgs phase (in which U(1) symmetry is spontaneously broken) but also the Coulomb phase (in which U(1) symmetry is unbroken).
In the Coulomb phase, the value of charged fields vanish.
In Ref.~\cite{Ferrara:2013rsa}, it was called de-Higgsed phase, and it was only recovered when $g=0$.
To enlarge the field space to include the Coulomb phase, we modify Eq.~\eqref{K_phi_1-inf},
\begin{align}
K= -3 \alpha \log \left[+ \log \biggl( \gamma+ |\Phi|^2 \biggr) \right] + 3 \beta \log \left( \gamma + |\Phi|^2 \right). \label{K_phi_1-inf_ex}
\end{align}
For example, consider the case $\gamma=1$.
Now, the field $\Phi$ is defined everywhere (including $\Phi=0$) as long as $|\Phi|$ is finite.
Inflationary dynamics is unchanged because inflation happens at $|\Phi|\gg 1$.
Alternatively, one has
\begin{align}
K= -3 \alpha \log \left[ -\log \left( \frac{ | \tilde{\Phi} |^2}{1+\gamma | \tilde{\Phi} |^2} \right) \right] - 3 \beta \log \left( \frac{| \tilde{\Phi} |^2}{1+\gamma | \tilde{\Phi} |^2 } \right), \label{K_phi_0-1_ex}
\end{align}
where $\gamma =1$ and $\tilde{\Phi}$ is defined everywhere except $\tilde{\Phi}=0$.

We can understand the validity of the above modification by looking back
the structure of the kinetic term \eqref{kin_pole} and the potential \eqref{pot_pole} for $\Phi$ before the modification, \textit{i.e.}~$\gamma=0$.
The kinetic term of $\Phi$ has a pole at the origin as well as $\tilde{\Phi}$, but inflation occurs at large $|\Phi|$.
Since the form of the kinetic term is the same as $\tilde{\Phi}$, and the potential is similar to that of $\tilde{\Phi}$, similar flattening happens also for $\Phi$.
Although we cannot extend the field domain of $\tilde{\Phi}$ to $\tilde{\Phi}=0$ without affecting the pole at $\tilde{\Phi}=0$, we can extend the field domain of $\Phi$ to $|\Phi|\leq 1$ by adding a parameter $\gamma$ as in Eq.~\eqref{K_phi_1-inf_ex} to remove the pole at $\Phi=0$,
since the pole at $\Phi=0$ does not do any essential role for flattening of the potential.
In this way, we can have an example of $D$-term attractor models which is also defined at the Coulomb phase.

For completeness, we introduce other fields to cancel gauge anomaly, and demonstrate stabilization of these fields during inflation.
We will rewrite $\Phi \to \Phi_{+}$ and introduce an oppositely charged field $\Phi_{-}$ and a neutral superfield $S$.
The stabilization of these additional fields turns out to be tricky, which will be discussed separately in Sec.~\ref{sec:stabilization}.

Now, let us discuss the properties of the generalized $D$-term attractor
models and their predictions of $(n_{\text{s}},r)$.  For
completeness, we also consider possibilities other than $\gamma=1$.  The
K\"ahler potential is given by
\begin{align}
	K = -3\alpha \log\left[ \log\biggl( \gamma + |\Phi_+|^2 \biggr) \right] + 3\beta \log\left( \gamma + |\Phi_+|^2 \right), \label{K_attractor}
\end{align}
for real and positive constants of $\alpha, \beta$, and a real constant $\gamma$.
We consider the region where $\gamma + |\Phi_{+}|^2 >1$ otherwise the K\"{a}hler potential~\eqref{K_attractor} becomes complex (except for even $\alpha$).
Writing $\varphi \equiv \sqrt{2}|\Phi_+|$, we obtain the following kinetic term and $D$-term potential:
\begin{align}
	&\mathcal L_{\text{K}} = \frac{3 \left[ \alpha \varphi^2/2 + \gamma \log\left( \gamma + \varphi^2/2 \right)\left( \beta\log\left( \gamma + \varphi^2/2 \right)-\alpha \right) \right]}
	{(\gamma+\varphi^2/2)^2 \left[ \log\left( \gamma + \varphi^2/2 \right) \right]^2}
	 \frac{\left(\partial_{\mu} \varphi \right)^2}{2}, \label{kin_att} \\
	 &V = \frac{9g^2}{2} \left(\frac{\varphi^2/2}{\gamma+\varphi^2/2} \right)^2\left(\beta - \frac{\alpha}{\log\left( \gamma + \varphi^2/2 \right)} \right)^2. \label{pot_att}
\end{align}
The potential becomes flat for $\varphi\to \infty$.
Although the kinetic term also approaches to $0$, the dependence on $\varphi$ is rather weak due to the term with $\alpha$,
and actually we can realize a successful inflation.
There are some noticeable features in this model.

\begin{itemize}
\item For $\gamma <0$, both the kinetic term and potential positively diverge at $\varphi=\varphi_p$ where
\begin{align}
	\varphi_p = \sqrt{ 2(1-\gamma) }.
\end{align}
The region of $\varphi < \varphi_p$ and $\varphi > \varphi_p$ are separated by the infinite potential barrier.  
The potential has a minimum with vanishing cosmological constant at $\varphi=\varphi_m$ where
\begin{align}
	\varphi_m = \sqrt{ 2\left(\exp\left(\frac{\alpha}{\beta}\right)-\gamma\right) }.   \label{phim}
\end{align}
Since both $\alpha$ and $\beta$ are positive, we always have $\varphi_m > \varphi _p$, and the potential becomes flat for $\varphi \to \infty$.  
Depending on parameters, there may be additional extremal points of the potential.
The equation which specifies such points is the same as one that determines the points where the kinetic term coefficient vanishes, namely
\begin{align}
 \alpha \varphi^2/2 + \gamma \log\left( \gamma + \varphi^2/2 \right)\left( \beta\log\left( \gamma + \varphi^2/2 \right)-\alpha \right) = 0. \label{extrema}
\end{align}
For sufficiently small $\beta$ for fixed $\alpha$ (or large $\alpha$ for fixed $\beta$), there are no solutions to the above equation in the region $\varphi>\varphi_p$.
Then, we do not have additional minima or maxima of the potential, and also the kinetic term has the physical sign.
In this case, slow-roll inflation is possible in the plateau region $\varphi \gg \varphi_m$ which is smoothly connected to the vacuum at $\varphi=\varphi_m$.

\item For $0 \leq \gamma < 1$, it is similar to the $\gamma<0$ case, but the kinetic term is positive definite.
There is a minimum of the potential at $\varphi=\varphi_m$ with $V=0$, and the potential asymptotes to a positive constant at the large field region. However, there are no additional extrema for any $\alpha$ and $\beta$ in this case.

\item For $\gamma=1$, 
$\varphi_p$ becomes $0$ and the potential at $\varphi=0$
has a finite value $V(0)=9g^2\alpha^2/2$. The kinetic term also remains finite at that point.
The potential minimum is still given by \eqref{phim}. This is a symmetry-breaking type potential.

\item For $1 <\gamma <\exp(\alpha/\beta)$, there are two potential minima
at $\varphi=0$ and $\varphi_m$ with $V=0$, which are separated by a finite potential barrier whose maximum is at $\varphi=\varphi_b<\varphi_m$.
Here, $\varphi_b$ is determined as a solution to Eq.~\eqref{extrema}. 
At this point, the kinetic term vanishes.
Since the kinetic term becomes negative for $\varphi<\varphi_b$, we must be careful about the dynamics after inflation.
Indeed, the height of the potential barrier satisfies $V(\varphi_b)<V(0)|_{\gamma=1}=9g^2\alpha^2/2$, which never exceeds
the vacuum energy during inflation $V(\varphi \to \infty) = 9g^2\beta^2/2$ if $\beta \ge \alpha$.
In such a case the inflaton field climbs over the potential barrier to enter the range $\varphi<\varphi_b$ after inflation,
taking a wrong sign of the kinetic term. This problem can be avoided if $\alpha>\beta$ and $\gamma$ is sufficiently close to $1$\footnote{
Alternatively, this may be solved by introducing an $F$-term potential for inflaton since it contains the inverse of K\"{a}hler metric, which becomes infinitely large as $\varphi \to \varphi_b$.
}.
 
\item For $\exp(\alpha/\beta) \leq \gamma$, the kinetic term is regular in all the field space and there is only one minimum of the potential at $\varphi=0$ with $V=0$.
In the large $\gamma$ limit, the potential in terms of the canonical field is approximated as that of the minimal model, $V\simeq g^2 \tilde{\varphi}^4 /8$, in the leading order of $1/\gamma$ expansion.

\end{itemize}

\begin{figure}[t]
\begin{center}
\includegraphics[width= 0.618034 \columnwidth]{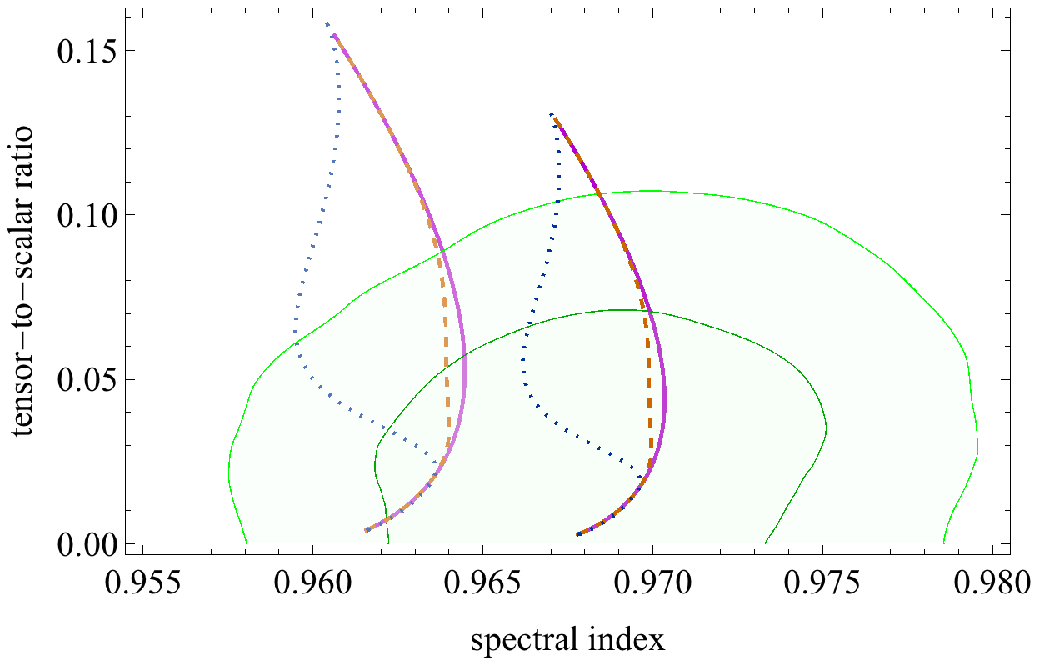}
\caption{
	$(n_{\text{s}},r)$ for $\gamma=0$ (purple solid), 0.1 (orange dashed), and 1 (blue dotted), where we have assumed $\alpha=\beta$.
	Lighter and darker lines correspond to the cases with $N_e = 50$ and $60$, respectively.
	For each line, we varied $\beta = 1~\textendash~10^5$.
	Green curves represent 68\% and 95\% confidence regions of the Planck TT+lowP+BKP+lensing+BAO+JLA+$H_0$ constraint (adopted from Fig.~21 of Ref.~\cite{Ade:2015xua}).
}
\label{fig:model2}
\end{center}
\end{figure}

 Using the method described in Appendix~\ref{sec:app}, we calculated the prediction of $(n_{\text{s}},r)$,
 which is shown in Fig.~\ref{fig:model2}.
 We see that the predicted values for $n_{\text{s}}$ and $r$ nicely fit the observational results, especially for $\alpha \lesssim 1$. 
 Let us discuss a limiting case with 
 $\varphi^2/2 \gg \gamma, \gamma\log(\varphi^2/2)$, and $(\gamma\beta /\alpha) \log^2(\varphi^2/2)$.
 In this limit the canonical field is given by
\begin{align}
	\tilde\varphi \simeq \int_{\varphi_m}^{\varphi} \frac{\sqrt{6\alpha}}{\varphi\log(\varphi^2/2) } d\varphi
	= \sqrt{\frac{3\alpha}{2}}\log \left[\frac{\log\left(\frac{\varphi}{\sqrt{2}}\right)}{\log\left(\frac{\varphi_m}{\sqrt{2}}\right)}\right].
\end{align}
In terms of the canonical field, the potential reduces to Eq.~\eqref{Vatt} and hence the same prediction~\eqref{att_ns-r} is recovered.  
 The field value of $\varphi$ at the $e$-folding number $N_e$ is given by
\begin{align}
	\varphi_{N_e} \simeq \sqrt{2}\exp\left(\frac{2N_e}{3\beta}\right).  \label{phi_N}
\end{align}
For consistency of the approximation, we should have 
\begin{align}
	\exp\left(\frac{4}{3\beta}\right) \gg \gamma,~\frac{\gamma}{\beta}~\text{and}~\frac{\gamma}{\alpha\beta}. \label{approximation}
\end{align}
When these conditions are not satisfied, \textit{e.g.} with large $\beta$ and nonzero $\gamma$, there are sizable corrections to the above potential.
An example of $(n_{\text{s}}, r)$ prediction for such a case with $\beta=\alpha$ is shown in Fig.~\ref{fig:model2}.

Let us comment on the meaning of the parameters and curvature of the K\"{a}hler manifold specified by eq.~\eqref{K_phi_1-inf_ex}.
$\gamma$ parametrizes deformation from the original $D$-term attractor model~\eqref{K_phi_1-inf}.
In the absence of $\gamma$, $\alpha$ defines the curvature of the K\"{a}hler manifold~\eqref{curvature}, and controls the way the inflaton is normalized.  This, in turn, determines the slope of the potential and hence the tensor-to-scalar ratio.  Meanwhile, $\beta$ controls the overall normalization of the potential.
In the presence of $\gamma$, the K\"{a}hler curvature now depends on the field value $\varphi$ as well as the parameters $\alpha$, $\beta$, and $\gamma$.
The exact expression is too long to be displayed here, but we can discuss some limits.
First of all, the limit $\gamma \to 0$ or $\varphi \to \infty$ reduces to the undeformed result, $R=-2/(3\alpha)$.
This reflects the fact that the effects of the deformation is minor in the large field limit.
The opposite limit $\gamma \to \infty$ leads to $R= 2/(3\beta)$.  However, in this limit, the tensor-to-scalar ratio is approximately
independent of $\beta$, as discussed above.
Thus, with general $\gamma$, the K\"{a}hler curvature is not directly
related to the inflationary observables.

%%%%%%%%%%%%%%%%%%%%%%%%%%%%%%%%%%%%%%%%%%%%%%%%%%
\subsection{Stabilization of other fields}
\label{sec:stabilization}
%%%%%%%%%%%%%%%%%%%%%%%%%%%%%%%%%%%%%%%%%%%%%%%%%%

As mentioned before, there are additional fields $\Phi_-$ and $S$, which must be stabilized during inflation.
Here, we consider the following superpotential,
\begin{align}
W = \lambda S \Phi_{+} \Phi_{-}. \label{W_attractor}
\end{align}
The masses of $\Phi_-$ and $S$ in the canonical basis during inflation are given by [see Eq.~\eqref{mass_F_model1}]
\begin{align}
	m_{\Phi_-,F}^2 = m_{S,F}^2 = e^K K_{S\bar S}^{-1} K_{\Phi_-\bar \Phi_-}^{-1}\frac{\lambda^2 \varphi^2}{2}.
\end{align}
If we naively extend the K\"ahler potential~\eqref{K_attractor} by introducing additional fields,
\begin{align}
K= -3 \alpha \log \left[ \log \biggl(\gamma+|\Phi_{+}|^2 + |\Phi_{-}|^2 +|S|^2 \biggr) \right] + 3 \beta \log \biggl( \gamma + |\Phi_{+}|^2 + |\Phi_{-}|^2+|S|^2 \biggr),
\label{K_attractor_stabilize}
\end{align}
masses of $\Phi_-$ and $S$ become
\begin{align}
	m_{S,F}^2 = m_{\Phi_-,F}^2 \simeq \left( \frac{\varphi^2}{2} \right)^{3(\beta+1)} \frac{\lambda^2}{9\beta^2} \left( \log\left( \frac{\varphi^2}{2} \right)\right)^{-3\alpha}.
\end{align}
Here and hereafter, our main interest is the attractor regime, so we have neglected $\gamma$ and assumed $\beta \log (\varphi^2/2) \gg \alpha$.
Note that in the non-canonical basis, $\varphi$ exponentially grows as $\exp(2N_e/3\beta)$ as seen from Eq.~\eqref{phi_N}.
Thus we find that
\begin{align}
	m_{S,F}^2 = m_{\Phi_-,F}^2 \propto \exp\left(
 \frac{4(\beta+1)}{\beta}N_e \right) \left(\frac{3\beta}{4N_e}\right)^{3\alpha}.
\end{align}
It means that their masses become at least $e^{{\cal O}(N_e)}$ times larger during the
last $N_e \simeq 60$ $e$-foldings, and that they easily exceed the
Planck scale.  This situation is similar to that of the polynomial
models described in Sec.~\ref{sec:D}.

The difficulty described above can be alleviated if we introduce the
following terms in addition to Eq.~\eqref{K_attractor_stabilize},
\begin{equation}
 h|S|^2\left(|\Phi_+|^{6\beta+4} + |\Phi_-|^{6\beta+4}\right). \label{K_attractor_stabilize_h}
\end{equation}
In this case, masses from $F$-term potential read
\begin{equation}
m_{S,F}^2 = m_{\Phi_-,F}^2 \simeq \frac{\lambda^2}{3\beta h}\left[\log\left(\frac{\varphi^2}{2}\right)\right]^{-3\alpha}. \label{mass_F}
\end{equation}
Note that the exponentially large factor $e^K K^{-1}_{\Phi_-\bar{\Phi}_-}\varphi^2 \propto \varphi^{6\beta+4}$ is canceled by
$K_{S\bar{S}}^{-1}\propto  \varphi^{-(6\beta+4)}$. In order to avoid the appearance of super-Planckian masses before the end of inflation,
we require the following condition,
\begin{equation}
\frac{\lambda^2}{3\beta h}\left(\frac{3\beta}{4}\right)^{3\alpha} < 1. \label{condition_Planck}
\end{equation}

It is also necessary to check that the $F$-term masses~\eqref{mass_F} actually stabilize $\Phi_-$ and $S$
against a negative contribution from $D$-term potential and the Hubble friction.
Due to the additional terms in Eqs.~\eqref{K_attractor_stabilize} and~\eqref{K_attractor_stabilize_h}, the $D$-term potential is modified as follows,
\begin{align}
V =& \frac{g^2}{2}\bigg[\frac{3(|\Phi_+|^2-|\Phi_-|^2)}{\gamma+|\Phi_+|^2 + |\Phi_-|^2+|S|^2}\left(\beta-\frac{\alpha}{\log(\gamma+|\Phi_+|^2 + |\Phi_-|^2+|S|^2)}\right) \bigg. \nonumber\\
&\qquad \bigg.+h(3\beta+2)|S|^2\left(|\Phi_+|^{6\beta+4} - |\Phi_-|^{6\beta+4}\right)\bigg]^2,
\end{align}
which leads to the following masses for $\Phi_-$ and $S$ in the canonical basis,
\begin{equation}
m_{\Phi_-,D}^2 \simeq -6g^2\beta \quad \text{and} \quad  m_{S,D}^2 \simeq 3g^2\beta(3\beta+2).
\end{equation}
Two fields are stabilized if the total masses $m_{\Phi_-,F}^2+m_{\Phi_-,D}^2$ and $m_{S,F}^2+m_{S,D}^2$
remain larger than $H^2 \simeq V/3 \simeq 3g^2\beta^2/2$ during inflation.
Since the negative contribution $m_{\Phi_-,D}^2$ dominates over $H^2$ for $\beta<4$,
it is sufficient to require that $m_{\Phi_-,F}^2 \gg -m_{\Phi_-,D}^2$.
From this condition, we find
\begin{equation}
\frac{\lambda^2}{3\beta h}\left(\frac{3\beta}{4 N_e}\right)^{3\alpha} \gg 6 g^2\beta. \label{condition_stabilization}
\end{equation}
Furthermore, some parameters are constrained by the requirement that they explain the amplitude of the observed power spectrum of 
the curvature perturbation~\cite{Ade:2015lrj},
\begin{equation}
\mathcal P_\zeta = \frac{V}{24\pi^2 \epsilon} \simeq \frac{9g^2\beta^4}{64\pi^2\alpha} \left[\log\left(\frac{\varphi^2}{2}\right)\right]^2 \simeq \frac{g^2\beta^2N_e^2}{4\pi^2\alpha} \simeq 2.2\times 10^{-9}.
\label{Ps_normalization}
\end{equation}
Combining Eqs.~\eqref{condition_Planck},~\eqref{condition_stabilization} and~\eqref{Ps_normalization}, we obtain
\begin{equation}
\frac{72\pi^2 \alpha A_s}{N_e^2}\left(\frac{4 N_e}{3\beta}\right)^{3\alpha} < \frac{\lambda^2}{h} < 3\beta \left(\frac{4}{3\beta}\right)^{3\alpha},
\label{constraint_lambda}
\end{equation}
where $A_s = 2.2\times 10^{-9}$.

The condition~\eqref{constraint_lambda} is easily satisfied for $\alpha\simeq \beta \ll 1$.
On the other hand, if the power exponent of $|\Phi_+|^2$ in Eq.~\eqref{K_attractor_stabilize_h} is some integer,
the value of $\beta$ is constrained, and its smallest value is $\beta = 1/3$.
In this case, larger values for $\alpha$ and/or smaller values for $g$ are required in order to satisfy the condition~\eqref{Ps_normalization},
but $\alpha$ cannot be arbitrarily large since it easily conflicts with the constraint~\eqref{constraint_lambda}.
Furthermore, a larger value of $\alpha$ leads to a large tensor to scalar ratio [see Eq.~\eqref{att_ns-r}],
which also conflicts with observational results.

%%%%%%%%%%%%%%%%%%%%%%%%%%%%%%%%%%%%%%%%%%%%%%%%%%
\section{K\"{a}hler potential for $D$-term pole inflation}
\label{sec:pole}
\setcounter{equation}{0}
%%%%%%%%%%%%%%%%%%%%%%%%%%%%%%%%%%%%%%%%%%%%%%%%%%
The $\alpha$-attractor is characterized by a second order pole in the coefficient of the inflaton kinetic term~\cite{Galante:2014ifa}.
Its prediction is $n_{\text{s}}=1 - 2/N_e$ and $r= 12 \alpha /N_e^2$.
This is generalized to poles of arbitrary order~\cite{Galante:2014ifa, Broy:2015qna, Terada:2016nqg}.
When the inflaton field $\tilde{C}$ has a pole of order $p$ in the kinetic term, we redefine the origin of the field so that the pole coincides with the origin. 
Then, we have the following inflaton Lagrangian,
\begin{align}
\left( \sqrt{-g} \right)^{-1} \mathcal{L}= \frac{1}{2} \frac{a_p}{\tilde{C}^p} \left( \partial_{\mu} \tilde{C} \right)^2 -V(\tilde{C}).
\end{align}
For the sign of the kinetic term to be physical, $a_p$ with even $p$ has to be positive.
For odd $p$, if $a_p$ is negative, we redefine $\tilde{C}\to - \tilde{C}'$ and $a_p \to - a_p'$, and drop the primes.
Thus, without loss of generality, we can assume $a_p>0$ and $\tilde{C}>0$.
The parameter $\alpha$ of the $\alpha$-attractor is related to $a_2$ as $a_2 = 3 \alpha / 2$.
If the potential has a positive finite value at its origin and is smoothly connected to a vacuum,\footnote{
For more general cases in which the potential also has a pole at $\tilde C=0$, the canonical potential becomes an exponential term (power-law inflation) or monomial (chaotic inflation)~\cite{Rinaldi:2015yoa, Terada:2016nqg}.
In this paper, we are mainly interested in plateau-type potentials, and do not consider singular potentials.
See Sec.~\ref{sec:symbrk}, however, for a pole with a small coefficient as a small shift-symmetry breaking effect.
}
the spectral index and tensor-to-scalar ratio are given by~\cite{Galante:2014ifa}
\begin{align}
n_{\text{s}}=& 1 - \frac{p}{(p-1)N_e}, &   r=& \frac{8 a'_p{}^{\frac{1}{p-1}}}{(p-1)^{\frac{p}{p-1}}N_e^{\frac{p}{p-1}}},
\end{align}
where $a_p$ is rescaled by the ratio of the constant and linear terms in $V(\tilde{C})$, $a'_p = a_p (-2a/b)^{p-2}$ (see Eq.~\eqref{condition_P} below where $a$ and $b$ are introduced).

In this section, we derive the form of K\"{a}hler potential leading to pole inflation of order $p$ in the case of $D$-term inflation.
In Ref.~\cite{Broy:2015qna}, the form of K\"{a}hler potential is studied for $p=2$ with higher order pole corrections.
We discuss general $p$, and derive the K\"{a}hler potential in terms of the ordinary U(1) charged field $\Phi$.
We investigate effects of higher order poles in the next subsection.

Using the general prescription given in Ref.~\cite{Ferrara:2013rsa}, we can construct any monotonically increasing potential.
The canonical inflaton potential for pole inflation is known~\cite{Broy:2015qna}, so we can construct K\"{a}hler potential leading to the potential.
Though equivalent, we focus in the following analysis more on the variable $\tilde{C}$ having a pole in the kinetic term because it is the essence of pole inflation.

To derive the conditions on the K\"{a}hler potential, it is useful to deal with a shift-symmetric field $\Lambda=\log(\Phi)$ because derivative of $K$ with respect to $\Lambda$ and its conjugate are same,
\begin{align}
K(\Lambda, \bar{\Lambda})=& K(\Lambda + \bar{\Lambda}), &   K_{\Lambda}=&K_{\bar{\Lambda}}. 
\end{align}
Similar relations hold for higher derivatives.
We therefore introduce a real variable $C= \Lambda + \bar{\Lambda}$ as in Ref.~\cite{Ferrara:2013rsa}, and the K\"{a}hler potential is effectively a function of a single variable, $K=K(C)$.

This $C$ does not necessarily coincide with $\tilde{C}$ introduced above, so $C=C(\tilde{C})$.
The coefficient of the kinetic term is given by the K\"{a}hler metric, so we require
\begin{align}
\frac{K''}{2} \left( \frac{\text{d}C}{\text{d}\tilde{C}} \right)^2 = \frac{ a_p}{\tilde{C}^p}, \label{condition_K}
\end{align}
where prime denotes differentiation with respect to $C$. Since $\Lambda$ transforms by a constant shift under U(1) transformation, the $D$-term potential is given by
\begin{align}
V= \frac{g^2}{2} \left( K' \right)^2.
\end{align}
Therefore, $K'$ should be a regular function of $\tilde{C}$ at the origin.
Expanding it up to a first order, we obtain
\begin{align}
K' =  b + a \tilde{C}. \label{condition_P}
\end{align}
The relative sign of $a$ and $b$ should be negative because otherwise inflaton $\tilde{C}$ keeps roll down to the origin, and inflation does not end. 
Combining Eqs.~\eqref{condition_K} and \eqref{condition_P}, we obtain
\begin{align}
K'= b+ a \left( -\frac{2 a_p}{(p-1) a C } \right)^{\frac{1}{p-1}},
\end{align}
where we neglect an integration constant which can be absorbed by the shift of $C$.
Integrating the above equation, we find the following form for the K\"{a}hler potential,
\begin{align}
K=& \begin{cases} K_0 + b C - 2a_2 \log |C| & (p=2), \\
K_0 + b C - \frac{p-1}{p-2}\left( \frac{2 a_p}{p-1} \right)^{\frac{1}{p-1}} \left( -aC\right)^{\frac{p-2}{p-1}}   &  (p \neq 2), 
\end{cases}
\end{align}
where $C=\Lambda+\bar{\Lambda}= \log|\Phi|^2$, and $K_0$ is an integration constant.
For $p=2$, we reproduce Eqs.~\eqref{K_phi_1-inf} or \eqref{K_phi_0-1} depending on the sign of $b$ (or $C$) with identification $K_0=0$, $b=\pm 3\beta$, and $a_2=3\alpha/2$.

Although we can define the transformation of $C$ or $\Phi$ corresponding to the shift of canonical inflaton $\tilde{\varphi}$, we do not find a simple formula or interpretation for the cases with $p \neq 2$.

As demonstrated in Sec.~\ref{sec:stab} and Sec.~\ref{sec:stabilization}, we can introduce fields for anomaly cancellation and stabilize them during inflation.  It is a straightforward generalization, so we do not repeat it here.

%%%%%%%%%%%%%%%%%%%%%%%%%%%%%%%%%%%%%%%%%%%%%%%%%
\subsection{Symmetry breaking effects} \label{sec:symbrk}
%%%%%%%%%%%%%%%%%%%%%%%%%%%%%%%%%%%%%%%%%%%%%%%%%
Next, we consider effects of higher order poles as symmetry breaking effects~\cite{Broy:2015qna},
\begin{align}
\left( \sqrt{-g} \right)^{-1} \mathcal{L}=\frac{1}{2} \left( \frac{a_p}{\tilde{C}^p} + \frac{a_q}{\tilde{C}^q} \right) \left( \partial_{\mu} \tilde{C} \right)^2 -V(\tilde{C}), \label{L_poles}
\end{align}
where $q>p$ without loss of generality, and we assume $ \frac{a_p}{\tilde{C}^p} \gg \frac{a_q}{\tilde{C}^q} $ during inflation because otherwise the effect of pole of order $p$ can be negligible and the previous discussion applies with substitution $p=q$.
We treat the $a_q$ term as a perturbation.
The observational effect on $(n_{\text{s}},\, r)$ due to this perturbation is given by the following shifts~\cite{Broy:2015qna}
\begin{align}
\delta n_{\text{s}} =& - \frac{(q-p)(q-p-1)(p-1)^{\frac{q-2p+1}{p-1}}a'_q}{a'_p{}^{\frac{q-1}{p-1}}(q-1)}N_e^{\frac{q-2p+1}{p-1}}, \\
\delta r=& - \frac{8(q-p-1)(p-1)^{\frac{q-2p}{p-1}}a'_q}{a'_p{}^{\frac{q-2}{p-1}}(q-1)}N_e^{\frac{q-2p}{p-1}},
\end{align}
where $a'_p = a_p (-2a/b)^{p-2}$ and $a'_q = a_q (-2a/b)^{q-2}$.

Due to the existence of higher order poles, Eq.~\eqref{condition_K} is replaced with
\begin{align}
\frac{K''}{2} \left( \frac{\text{d}C}{\text{d}\tilde{C}} \right)^2 = \frac{a_p}{\tilde{C}^p} +  \frac{a_q}{\tilde{C}^q}.
\end{align}
Up to the first order in $a_q$, the additional term in the K\"{a}hler potential is
\begin{align}
\Delta K = \frac{-(p-1)a_q}{(q-1)(q-2)a_p} \left( \frac{2 a_p}{p-1} \right)^{\frac{p-q+1}{p-1}}\left(-aC\right)^{\frac{q-2}{p-1}}. \label{DeltaK}
\end{align}
In the case of $\alpha$-attractor ($p=2$), this reduces to
\begin{align}
\Delta K = \frac{-2 a_q}{(q-1)(q-2)} \left(\frac{-aC}{3 \alpha}\right)^{q-2}.
\end{align}
Thus, higher order terms with respect to $C=\Lambda+\bar{\Lambda}= \log|\Phi|^2$ in the K\"{a}hler potential can be interpreted as symmetry breaking terms.

One may also consider a pole in the potential suppressed by a small number as another source of shift symmetry breaking.
In this case, Eq.~\eqref{condition_P} is replaced with
\begin{align}
K'= \frac{c}{\tilde{C}^s} + b + a \tilde{C}, \label{condition_P2}
\end{align}
where we assume the first term is subdominant to the second term during observable length of inflation.
This leads to $s$-th and $(s-1)$-th order poles in the potential at the first order of $c$.
The latter is subdominant assuming the term $b$ is dominant in Eq.~\eqref{condition_P2} during inflation.
Such a term affects $n_{\text{s}}$ and $r$~\cite{Terada:2016nqg},
\begin{align}
\delta n_{\text{s}} =& \frac{s(s+1)(p+2s) b_s }{(p+s)a'_p} \left( \frac{p-1}{a'_p} N_e \right)^{\frac{s-p+2}{p-1}} , \\
\delta r = & \frac{8s(p+2s)b_s}{(p+s)a'_p}  \left( \frac{p-1}{a'_p} N_e \right)^{\frac{s-p+1}{p-1}},
\end{align}
where $b_s = (2c/b) \times (-2a/b)^s$ measures the relative magnitude of the singular term in the potential. 
Combining Eq.~\eqref{condition_P2} with Eq.~\eqref{condition_K}, we can similarly obtain the correction to the K\"{a}hler potential up to a first order in $c$,
\begin{align}
\Delta K = -\frac{2c (p+2s) a_p}{a(p+s)(p+s-1)} \left( \frac{-(p-1)aC}{2a_p} \right)^{\frac{p+s-1}{p-1}}.
\end{align}
In the case of $\alpha$-attractor ($p=2$), this reduces to
\begin{align}
\Delta K = - \frac{6 c \alpha }{a(s+2)} \left( \frac{-aC}{3\alpha} \right)^{s+1} .
\end{align}
These are also higher order terms in $C$.
The similarity to the case of a higher order pole in the kinetic term can be understood as follows.
In Eq.~\eqref{condition_P2}, one can redefine the field as $a \tilde{C}' = a \tilde{C}+c \tilde{C}^{-s}$.
This eliminates the pole in the potential, but it induces a pole in the kinetic term.
Thus, to the first order of $c$, one can map the Lagrangian to the form of Eq.~\eqref{L_poles}.
In this way, small perturbation as a form of higher order terms $\propto C^{n}$ $(n>1)$ in the K\"{a}hler potential are equivalent to the shift symmetry breaking effect in terms of the canonical inflaton field $\tilde{\varphi}$, which can be interpreted either as an additional pole in the kinetic term or as a pole in the potential in terms of an intermediate field $\tilde{C}$.

%%%%%%%%%%%%%%%%%%%%%%%%%%%%%%%%%%%%%%%%%%%%
\section{Conclusion and discussions}
\label{sec:summary}
\setcounter{equation}{0}
%%%%%%%%%%%%%%%%%%%%%%%%%%%%%%%%%%%%%%%%%%%%

In this paper we have tried to construct a concrete model of
$D$-term inflation based on attractor models. 
First, we have shown that simple models of $D$-term
chaotic inflation do not fit the current data, unfortunately. In
addition, we have pointed out that the effective masses of $S$ and $\Phi_-$ during inflation is exponentially large and typically beyond the
Planck scale, which might destroy the validity of the calculations due
to quantum gravity effects. As shown explicitly, higher order corrections to
the K\"ahler potential improve the fit to the data and succeed in
suppressing masses of the additional fields adequately. However, without
symmetry reason, these corrections are uncontrollable.

Then, we have revisited the $D$-term inflationary attractor models in
this paper. These attractor models can be realized in the context of
pole inflation with a second order pole in the kinetic term. However, we
have seen the limitation on the structure of K\"{a}hler potential for
$D$-term inflationary attractor models. Actually, it is not
automatically guaranteed that a sufficiently flat potential is obtained
even if one specifies a K\"{a}hler potential which leads to a second
order pole in the kinetic term. This is because the $D$-term potential
is also determined by the K\"{a}hler potential. We have extended the
models in order to construct a workable example, in which K\"{a}hler
potential is defined in Coulomb phase as well as in Higgs phase, gauge
anomaly is cancelled and other required fields including the stabilizer
field are stabilized during inflation. These can be viewed as a first
step of UV completing the $D$-term attractor models.  For these models
with plateau type potentials, understanding the origin of shift symmetry
for canonical inflaton is important. We pointed out that it is a
symmetry under the transformation $\Phi = \Phi'{}^{\hat{c}}$ for the
$\alpha$-attractor (pole inflation with $p=2$) case.

Finally let us discuss the reheating of the universe in our model~\eqref{K_attractor}.  In
particular, let us consider the case of $\gamma < 1$, in which the
global potential minimum is given by $\varphi=\varphi_m$ (\ref{phim})
and $S=\Phi_-=0$.\footnote{At the potential minimum, the K\"ahler metric
$K_{\Phi_-\bar\Phi_-}$ becomes zero and the physical mass of $S$
diverges.  To avoid this, we may introduce a small correction, e.g., $K
\sim \zeta |\Phi_+|^2 |\Phi_-|^2$.  Although it can mildly affect the
K\"ahler potential \eqref{K_attractor_stabilize_h} required for safely
stabilizing the inflationary path, the qualitative discussion remains
intact.}  After inflation, inflaton begins to oscillate around the
minimum of the potential.  The inflaton and gauge boson masses around
the potential minimum are given by
\begin{align}
	m_\varphi^2 = m_{A}^2 = \frac{6g^2\beta^2}{\alpha}\left( \frac{\varphi_m^2/2}{\gamma+\varphi_m^2/2} \right)^2.
\end{align}
Since SUSY is preserved at the minimum, the masses of the inflaton and
the gauge boson, which are both members of a massive gauge
supermultiplet, are same.  Inflaton does not decay into gauge bosons or
gauginos for the kinematical reason.  To reheat the standard model
sector, we may introduce a (small) kinetic mixing between the gauge
bosons of the U(1) and the standard model hypercharge U(1)$_Y$.  Then
the inflaton or hidden gauge bosons decay into the standard model
U(1)$_Y$ gauge boson pair through the mixing.
The case of $\gamma=1$ is similar, but the cosmic string may be formed.
When $\gamma \geq \exp (\alpha / \beta )$, the inflaton is massless at the vacuum, and reheating becomes nontrivial.

%%%%%%%%%%%%%%%%%%%%%%%%%%%%%%%%%%%%%%%%%%%%
\section*{Acknowledgments}
%%%%%%%%%%%%%%%%%%%%%%%%%%%%%%%%%%%%%%%%%%%%

We would like to thank J.\ Yokoyama for his collaboration and
useful comments at the initial stage of the present work.  This work was
supported by the Grant-in-Aid for Scientific Research on Scientific
Research (No.~26$\cdot$10619 [TT]), Scientific Research A (No.26247042
[KN]), Young Scientists B (No.26800121 [KN]), Innovative Areas
(No.26104009 [KN], No.15H05888 [KN,~MY]), Scientific Research Nos.\
25287054[MY], 26610062[MY].  KS and TT are also supported by the
Grant-in-Aid for JSPS Fellows.

\appendix

%%%%%%%%%%%%%%%%%%%%%%%%%%%%%%%%%%%%%%%%%%%%%%%%%%
\section{Slow-roll inflation in non-canonical basis}
\label{sec:app}
\setcounter{equation}{0}
%%%%%%%%%%%%%%%%%%%%%%%%%%%%%%%%%%%%%%%%%%%%%%%%%%

Let us consider the Lagrangian with a non-canonical scalar field
\begin{align}
	\left( \sqrt{-g} \right)^{-1}\mathcal L = \frac{1}{2}f(\varphi)(\partial_{\mu} \varphi)^2 - V(\varphi).
\end{align}
The canonical field is
\begin{align}
	\tilde\varphi = \int \sqrt{f(\varphi)}d\varphi,
\end{align}
but often this integral cannot be analytically performed.
Thus it may be convenient to consider the inflaton dynamics in a non-canonical basis.

The equation of motion is
\begin{align}
	f \ddot\varphi + \frac{\dot f \dot\varphi}{2} + 3H f \dot\varphi + V_\varphi = 0.
\end{align}
The subscript $\varphi$ denotes the derivative with respect to $\varphi$.
The Friedmann equation is
\begin{align}
	3M_P^2 H^2 = \frac{f\dot \varphi^2}{2} + V.
\end{align}

In the slow-roll limit, we have
\begin{align}
	&3H f \dot\varphi + V_\varphi = 0,\\
	&3M_P^2 H^2 =V.
\end{align}
The slow-roll consistency conditions are
\begin{align}
	\epsilon \ll 1,~~\left|\eta - \frac{\xi}{2}\right|\ll 1,
\end{align}
where
\begin{align}
	\epsilon \equiv \frac{M_P^2}{2f}\left( \frac{V_\varphi}{V} \right)^2,~~~\eta \equiv \frac{M_P^2}{f}\frac{V_{\varphi\varphi}}{V},~~~
	\xi \equiv \frac{M_P^2 f_\varphi}{f^2}\frac{V_\varphi}{V}.
\end{align}
The slow-roll equation is rewritten in terms of the $e$-folding $N_e$ as
\begin{align}
	\frac{d\varphi}{dN_e} = \frac{V_\varphi}{3H^2 f},
\end{align}
so $N_e$ is calculated as
\begin{align}
N_e = \int _{\min [ \varphi_{N_e} , \varphi_{\text{end}}]}^{\max [ \varphi_{{N_e}}, \varphi_{\text{end}} ]} \sqrt{\frac{f}{2\epsilon}} \text{d}\varphi,
\end{align}
where $\varphi_{\text{end}}$ and $\varphi_{N_e}$ are the field values at the end of inflation and at $N_e$ $e$-foldings before that.
According to the $\delta N$ formalism~\cite{Starobinsky:1986fxa, *Sasaki:1995aw, *Sasaki:1998ug, *Lyth:2004gb}, the curvature perturbation is evaluated as
\begin{align}
	\zeta(\vec x) = \Delta N_e (\vec x) = \frac{H \delta\tilde\varphi (\vec x)}{\dot{\tilde\varphi}}.
\end{align}
Note that it is the canonical field $\delta\tilde\varphi$, not $\delta\varphi$, that obtains long wave quantum fluctuation of $ H_{\rm inf}/(2\pi)$
during inflation:
\begin{align}
	\langle \delta\tilde\varphi_k \delta\tilde\varphi_{k'}  \rangle = (2\pi)^3\delta(\vec k+\vec k') \frac{2\pi^2}{k^3}\mathcal P_{\delta\tilde\varphi},~~~
	\mathcal P_{\delta\tilde\varphi}= \left( \frac{H_{\rm inf}}{2\pi} \right)^2.
\end{align}
Then we obtain the dimensionless power spectrum of the curvature perturbation as
\begin{align}
	\mathcal P_\zeta = \left( \frac{H_{\rm inf}}{\dot{ \tilde\varphi}} \right)^2\mathcal P_{\delta\tilde\varphi} = \frac{f}{12\pi^2}\frac{V^3}{M_P^6 V_\varphi^2} = \frac{V}{24\pi^2 M_P^4 \epsilon}.
\end{align}
The scalar spectral index is given by
\begin{align}
	n_s-1 = \frac{d \ln \mathcal P_\zeta}{d \ln k} = \frac{d\varphi}{dN_e}\frac{d \ln \mathcal P_\zeta}{d \varphi}
	 = - 6\epsilon + 2\eta - \xi.
\end{align}
The tensor-to-scalar ratio is given by
\begin{align}
	r = 16\epsilon.
\end{align}
%%

%%%%%%%%%%%%%%%%%%%%%%%%%%%%%%%%%%%%%%%%%%%%%%%%%%
\section{Generic scalar potential in $D$-term inflation} \label{app:generic}
\setcounter{equation}{0}
%%%%%%%%%%%%%%%%%%%%%%%%%%%%%%%%%%%%%%%%%%%%%%%%%%

In this section we explicitly construct a ``generic scalar potential'' as an explicit realization of Ref.~\cite{Ferrara:2013rsa}
in terms of Higgs fields $\Phi_+$ and $\Phi_-$.
In contrast to the original model, it is well-defined at $\Phi_+=0$ and $\Phi_-=0$.
Let us consider the following K\"ahler potential
\begin{align}
	K = \sum_{n} \frac{c_n}{n} \left[ \ln \left(1+b_n|\Phi_+|^2 \right) \right]^n.
\end{align}
In the small field limit $\Phi_+ \to 0$, this is expanded as $K \sim |\Phi_+|^2 + ({\rm const}) \times |\Phi_+|^4 + \dots$,
but in the large field limit the behavior is significantly modified.
From this K\"ahler potential we obtain
\begin{align}
	&K_{\Phi_+} = \sum_{n} c_n \left[ \ln \left(1+b_n|\Phi_+|^2 \right) \right]^{n-1} \frac{b_n\Phi_+^*}{1+b_n|\Phi_+|^2},\\
	&K_{\Phi_+ \bar \Phi_+}=\sum_{n} c_n \left[ \ln \left(1+b_n|\Phi_+|^2 \right) \right]^{n-2}
	\frac{(n-1)b_n^2|\Phi_+|^2 + b_n \ln \left(1+b_n|\Phi_+|^2 \right) }{(1+b_n|\Phi_+|^2)^2}.
\end{align}
Note that the K\"ahler metric is regular at $\Phi_+ \to 0$ for $c_1\neq 0$.
The fields are canonically normalized in the limit $\Phi_+\to 0$ for $c_1 = 1/b_1$.
The $D$-term potential for $\Phi_+$ with $\Phi_-=0$ is given by
\begin{align}
	V_D = \frac{g^2}{2}\left (   \sum_{n} c_n \left[ \ln \left(1+b_n|\Phi_+|^2 \right) \right]^{n-1} \frac{b_n|\Phi_+|^2}{1+b_n|\Phi_+|^2}  \right)^2.
	\label{VD_m3}
\end{align}

Let us consider the limit $b_n|\Phi_+|^2 \gg 1$.
If $c_n$ $(n\geq 2)$ are non-zero, it is easy to see that the canonical field becomes $\tilde\varphi \sim \ln \varphi$.
Actually for $c_2 \gg |c_3|, |c_4|,\dots$, the kinetic term becomes
\begin{align}
	\mathcal L_{\text{K}} \simeq c_2 \frac{(\partial_{\mu} \varphi)^2}{\varphi^2}.
\end{align}
Therefore we have $\tilde\varphi \simeq \sqrt{2c_2}\ln \varphi$.
Then from Eq.~\eqref{VD_m3} we obtain the $D$-term potential just as polynomial of $\tilde\varphi$, 
\begin{align}
	V \simeq c_2 g^2 \tilde\varphi^2 \left(1 + \frac{\sqrt{2} c_3}{(c_2)^{3/2}}\tilde\varphi + \frac{2c_4}{c_2^2}\tilde\varphi^2 \right)^2,
\end{align}
where we assumed $c_2 \gg |c_3| \gg |c_4|$ and neglected $c_n$ with $n=1$ and $n\geq 5$.
Thus we obtain a polynomial type potential.
It is obvious that we can also obtain more general form by choosing $c_n$ appropriately.

\begin{figure}[t]
\begin{center}
\includegraphics[width= 0.618034 \columnwidth]{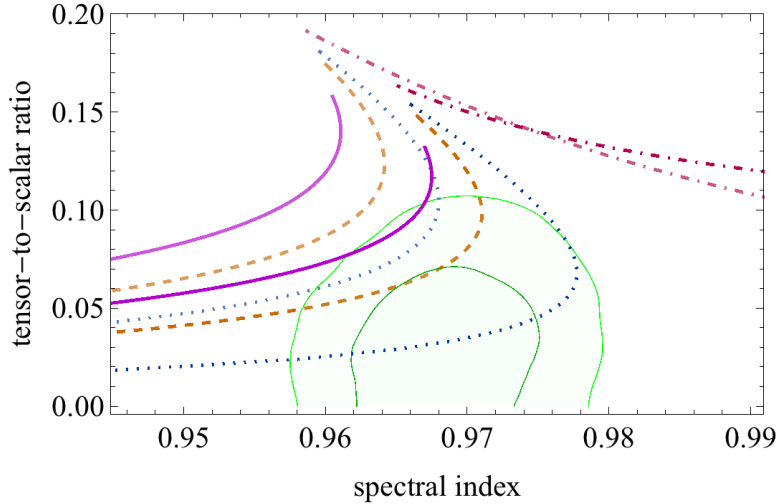}
\caption{
	$(n_s,r)$ for $c_4=0$ (purple solid), $2\times10^{-4}$ (orange dashed), $3\times10^{-4}$ (blue dotted), and $5\times10^{-4}$ (red dot-dashed). For each line, we varied $-c_3=0-4\times 10^{-2}$.
	Green curves represent 68\% and 95\% confidence regions of the Planck TT+lowP+BKP+lensing+BAO+JLA+$H_0$ constraint (adopted from Fig.~21 of Ref.~\cite{Ade:2015xua}).
}
\label{fig:model3}
\end{center}
\end{figure}

To illustrate impacts of this type of model,
we solved the dynamics of $\varphi=\sqrt{2}|\Phi_+|$ assuming $\Phi_-=0$.
For numerical calculation we take $b_n = b = 10$, $c_2 = 1$ and $g=1$.
The result is shown in Fig.~\ref{fig:model3} for $c_4=0, 2\times10^{-4}, 3\times10^{-4}$, and $5\times10^{-4}$. 
For each line, we varied $-c_3=0-4\times 10^{-2}$ within the range satisfying $V_\varphi > 0$ for the whole last 60 $e$-foldings.

The stability of $S$ and $\Phi_-$ during inflation is a nontrivial issue also in this model, 
but we can arrange the superpotential and K\"ahler potential so that they are stabilized at the origin while avoiding too large masses
in a similar fashion to that of Sec.~\ref{sec:stab}.

%%%%%%%%%%%%%%%%%%%%%%%%%%%%%%%%%%%%%%%%%%%%%%%%%%

\bibliographystyle{utphys}
\bibliography{ref.bib}
\end{document}